\documentclass{mn2e}

\title[Efficient gravitational lens surveys]
{Improving efficiency in radio surveys for gravitational lenses\\}
\author[N. Jackson and I.W.A. Browne]
{N. Jackson and I.W.A. Browne\\
The University of Manchester, Jodrell~Bank~Observatory, Macclesfield, Cheshire, SK11 9DL U.K.\\}

\begin{document}
\input{psfig}
\label{firstpage}
\maketitle

\begin{abstract}

Many lens surveys have hitherto used observations of large samples of
background sources to select the small minority which are multiply
imaged by lensing galaxies along the line of sight. Recently 
surveys such as SLACS and OLS have improved the efficiency of surveys by
pre-selecting double-redshift systems from SDSS. We explore other ways
to improve survey efficiency by optimum use of astrometric and
morphological information in existing large-scale
optical and radio surveys. The method exploits the small position
differences between FIRST radio positions of lensed images and the SDSS lens
galaxy positions, together with the marginal resolution of some larger 
gravitational lens systems by the FIRST beam.
We present results of a small pilot study with the VLA and MERLIN, and discuss the
desirable criteria for future surveys.

\end{abstract}


\section{Introduction}

Over the 30 years since the discovery of the first gravitational lens
system (Walsh et al. 1979) about 100 such systems have been discovered. 
The deflection of light in these systems can be used to probe the 
gravitational potential of the lensing galaxy. They can therefore be used
to investigate mass distributions in galaxies in the 0.5--20kpc region 
where the images form. This is the interesting region where the
transition between the dark matter-dominated halo and baryon-dominated
core occurs. Combined with optical spectroscopy, this is the only 
unambiguous measurement of galactic mass available at
cosmologically interesting distances. There have been numerous studies
of mass distributions in galaxies, particularly in cases where the system
gives enough constraints for a good lens model or when optical spectroscopy is also
available or both (e.g. Kochanek et al. 2001; Saha \& Williams 2001; Cohn et al.
2001; Mu\~noz et al. 2003; Treu \& Koopmans 2004; Dye \& Warren 2005, 
Koopmans et al. 2006). A further use of galaxy lensing 
is the calculation of the Hubble
constant using the time delay method of Refsdal (1964); the $H_0$ value
and error budget of this determination is somewhat controversial (e.g. Schechter
et al. 1997; Impey et al. 1998; Saha \& Williams 2001; Kochanek 2002;
Koopmans et al. 2003) but it nevertheless represents an
important, intrinsically clean method of $H_0$ determination on
cosmological scales.

Most early lens searches relied on systematic surveys of background
objects in order to pick out the few cases with galaxies close enough
to the line of sight to produce multiple imaging. In the radio, the
MIT-Greenbank survey (Bennett et al. 1986; Lawrence et al. 1986;
Hewitt et al. 1988) found five systems, a survey for lensed radio
lobes by Leh\'ar et al.  (2001) and Haarsma et al. (2005) found a few
further lens systems and the JVAS/CLASS surveys (Patnaik et al. 1992;
King et al. 1999; Myers et al. 2003; Browne et al. 2003) found 22 lens
systems. In a subsequent survey using a similar approach to CLASS
(Winn et al. 2001; Winn et al. 2002) found at least three lens systems
at Southern declinations.  Optical searches have included systematic
ground-based studies of optically-selected quasars (Crampton, McClure \&
Fletcher 1992; Surdej et al. 1993; Jaunsen et al. 1995; Kochanek, Falco \&
Schild 1995) and similar studies with the HST (Bahcall et al. 1992;
Maoz et al. 1992). More recently, the Sloan Digital Sky Survey has
been used for further lens searches (e.g. Inada et al. 2003). Oguri
et al. (2006) describe an algorithm based solely on the SDSS using
morphological and colour selection, developing earlier work by Pindor
et al. (2003). Pindor et al. use morphological selection by comparing
the $\chi^2$ of single-component and multi-component PSF fits to SDSS
images, and claim to be able to detect lens systems with separations
as small as 0\farcs7 and with flux ratios as much as 10:1.

Most searches are inefficient because typically many hundreds of
background sources must be searched in order to find one gravitational
lens, and it can be difficult to find an efficient filter which will
select all the lenses while rejecting false positives. The exception
is the Haarsma et al. (2005) survey which is discussed below, and,
increasingly, the SDSS morphological searches.  More recently,
however, major progress is being made using the redshift information
in the SDSS to pre-select systems with two discordant redshifts. The
major surveys to exploit this are the OLS survey (Willis et al. 2006)
and the SLACS survey (Bolton et al. 2006; Treu et al. 2006; Koopmans
et al. 2006) which have already found 24 systems between them and are
likely to discover between 50 and 100 lens systems. These are nearly
all radio-quiet systems with extended optical background sources which
are useful for lens modelling.

In this paper we address the question of making surveys for lensed
radio sources more efficient where detailed redshift information is
not available. There are many surveys of objects, such as the FIRST
survey, which in principle contain several thousand objects in which
the effects of lensing should be visible, but in which a direct search
of all images at high enough resolution to detect lenses would be
prohibitively expensive. This is the case even with the next
generation of telescopes such as EVLA and e-MERLIN. To illustrate the
extent of the problem, consider a survey of faint compact
flat-spectrum radio objects. All flat-spectrum objects above 30mJy
have already been searched for evidence of lensing in JVAS/CLASS, and
to obtain $\sim$10 times more objects, one would need to go a factor
of 10 fainter.  The increased sensitivity of future instruments allows
for this factor of 10, but to get a factor of 10 more lenses, either
10 times more targets need to be observed (i.e a factor of ten more
observing time is needed) or surveys will have to get a factor 10 more
efficient.

In this paper we explore ways of increasing efficiency of surveys for
lens systems targeted at compact radio sources. The method attempts to
make optimum use of existing general-purpose surveys. We describe the
results of a small pilot of an ``efficient'' survey, calculate the
expected lensing yield in a realistic situation, and consider the
prospects for larger surveys of this type in the future.

\section{Observational methods for finding radio lenses}

\subsection{Outline of the method}

As inputs we start with an optical and a radio survey of a patch of sky,
neither of which have the resolution on their own to find gravitational
lens system. For the purposes of the current investigation 
we distinguish five types of lens system:

\begin{enumerate}
\item Compact radio sources with images visible on the optical survey but 
with lens galaxies fainter than the optical survey limit
\item Compact radio sources with images below the optical survey limit but
with lens galaxies brighter than the optical survey limit
\item Compact radio sources where both the optical images and the lensing
galaxies are visible on the optical survey
\item Moderately compact, but steep-spectrum, isolated radio sources
with lens galaxies visible on the optical survey
\item Extended radio lobes associated with multiple sources, 
lensed by galaxies visible on the optical survey
\end{enumerate}

We note that the Haarsma et al. (2005) survey constitutes an attempt to
find lenses of the fifth type, and therefore do not consider them
further. In cases where the radio source
is compact we also need to consider the possibility separately of
detecting 2-image and 4-image systems.
We also note that, if the SDSS is used as the optical survey, 
JVAS/CLASS lenses would be classified as a mixture of types (ii) and (iii). 
Type (iii) cases are dominated by quasars from the strong-source
end of the survey; at the fainter end, radio sources with flux densities
$<$100~mJy are optically much fainter. In these cases, finding
redshifts for them can be difficult even with large telescopes (Marlow et al.
2000; Falco, Kochanek \& Mu\~noz 1998).

There are two potentially useful effects that may be seen, despite
limited resolution in existing radio and optical catalogues, due to
the fact that the radio catalogues are more sensitive to the source
brightness than the lens brightness.  First, lensing involves a
displacement of the centroid of bright images away from the lens. This
is particularly marked in the case of double-image systems, where the
brighter image is of order 1.5--2 Einstein radii from the lens, but is
also true for most four-image systems where the pair of merging
images, which dominates the source brightness, is located one Einstein
radius from the lens. We refer to this effect as the ``optical--radio
offset'' (Fig. 1). In the most optimistic case, assuming that the
optical image is dominated by light from the lens and not the lensed
images, we can detect this offset at levels well below the resolution
of both surveys. For instance, if FIRST is used as the radio survey,
displacement can be detected at the 3-$\sigma$ level down to
separations of approximately 1$^{\prime\prime}$ despite the
5$^{\prime\prime}$ resolution of FIRST. A test based on the
optical--radio offset becomes more powerful the more asymmetric the
lens system.

Second, the stretching introduced by the geometry of the multiple
imaging can be detected even if the image separation is below the formal
resolution, given sufficient signal-to-noise ratio.
In the case of double-image systems, the stretching is parallel to the vector
joining the centroid of the radio and optical source positions, and in
the case of quad lenses it is perpendicular to it. We refer to this as
the ``misalignment angle''. Conveniently, at least for double-image lens
systems, a test based on the misalignment angle becomes more powerful the
less asymmetric the system.

\begin{figure}
\psfig{figure=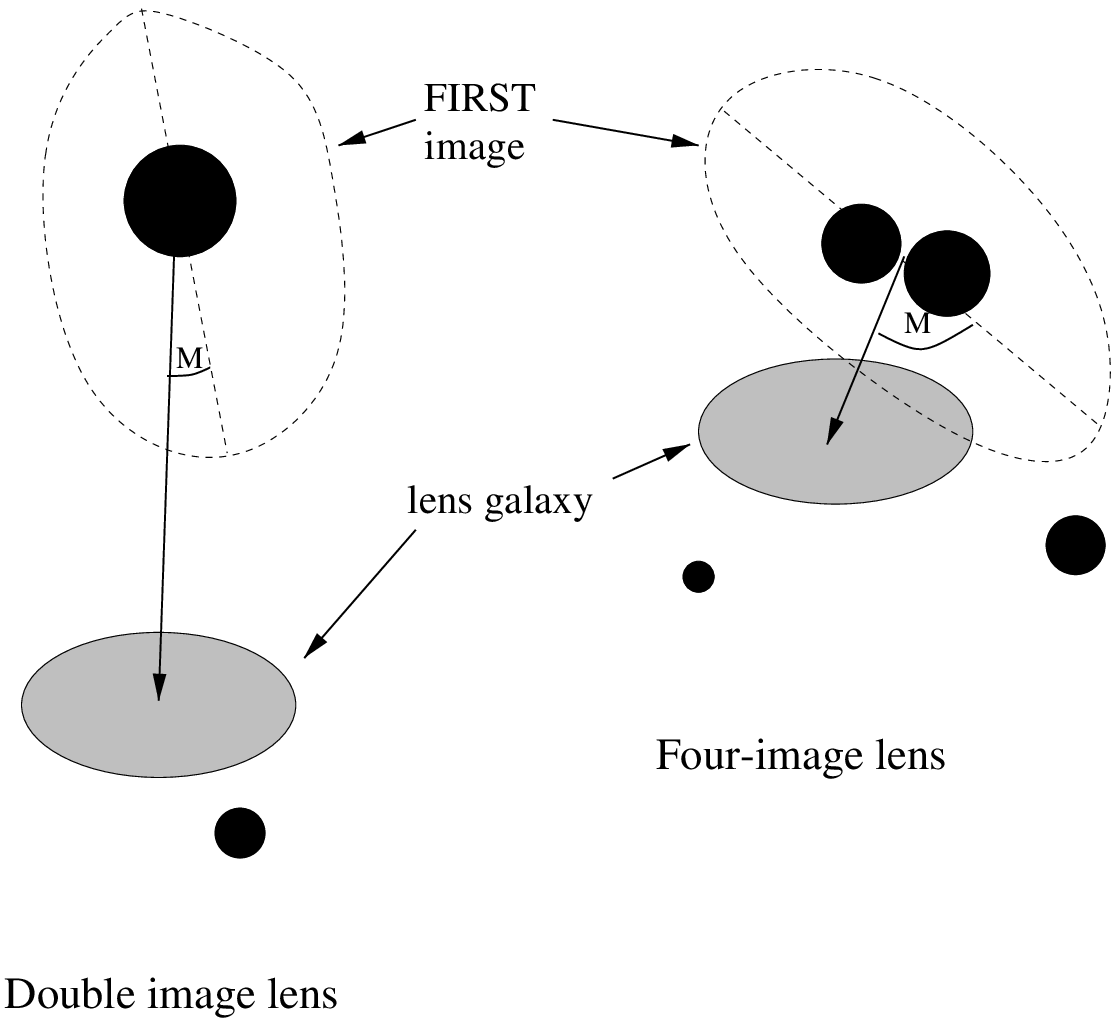,width=14cm}
\caption{Diagram of the expected configuration
for quad and double lens systems. In the case 
of double images, the FIRST centroid (dotted)
is expected to be offset from the lens galaxy and
to be extended roughly along the direction of the
optical-radio separation due to the influence of the
radio secondary (i.e. M=0). For quad lenses, the
FIRST radio and SDSS optical galaxy will also be
offset, because the close pair of images are very
bright, and the radio axis will in general be nearly
orthogonal to the optical--radio offset.}
\end{figure}

Figure 1 illustrates the overall geometry of these two situations. The
misalignment angle $M$ is defined as the angle between the radio
component major axis and the line joining the radio and optical
centroids, and should be close to 0$^{\circ}$ for double lenses and
close to 90$^{\circ}$ for quads. Thus lensed systems should be
predominantly located in certain defined regions of the
M/optical--radio offset plane. This is the foundation of the method we
are proposing. It should be emphasised, however, that although this
method at first glance gives powerful detection of lens systems with a
minimum of false positives, Figure 1 represents an ideal
situation. When there are systematic astrometric errors within the
surveys, or for type (iv) lenses with some extended structure, the
situation is more complicated and it is important to use simulations
and pilot observations to investigate the effectiveness of such
methods in more detail. As a preliminary illustration that this method
can work, we plot in Figure 2 some actual CLASS data on three
gravitational lens systems. It can be seen that, provided the lensed
images are not optically dominant, the optical--radio offset can be
used to distinguish cases of lensing. In practice, the chances of the
lensed object being optically bright diminish hugely with decreasing
radio flux density (Falco, Kochanek \& Mu\~noz 1998; Marlow et al. 
2000) and thus the test should become more efficient for fainter radio 
sources. The test is also particularly helpful for very asymmetric
lenses, which are the best candidates for searching for central images
to give constraints on the mass distribution in the centre of the
lensing galaxy (e.g. Boyce et al. 2006).

\begin{figure*}
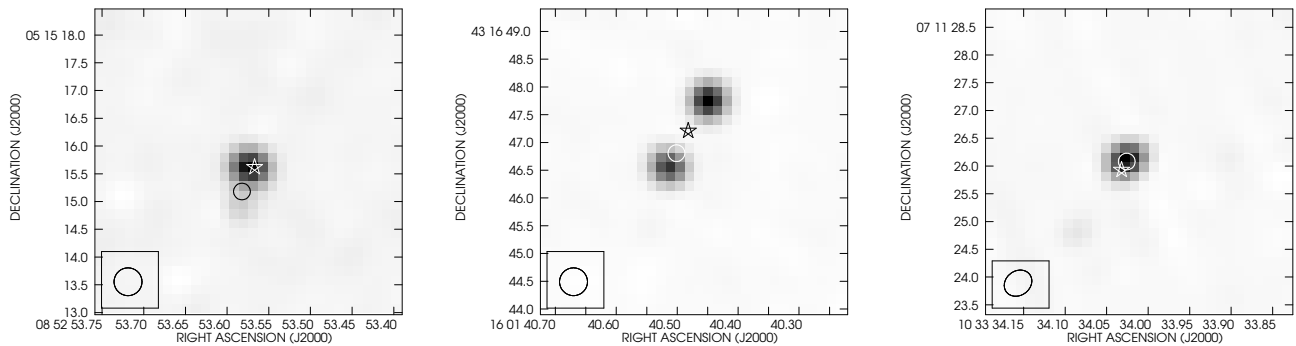

\begin{tabular}{ccc}
\psfig{figure=P0850,width=5.5cm,angle=-90}&
\psfig{figure=P1600,width=5.5cm,angle=-90}&
\psfig{figure=P1030,width=5.5cm,angle=-90}\\
\end{tabular}
\caption{Actual data for three CLASS lenses: B0850+054 (left), B1600+434
(centre) and B1030+074 (right), mapped using a restoring beam similar to the
B-array 8.4-GHz observations used in this paper. The FIRST and SDSS
positions are marked with a star and circle, respectively. Note that in
the first two cases, the SDSS position is offset from the radio peak,
reflecting the fact that the galaxy is offset from the brightest lensed
image as in Figure 1. Note that this method does not work for B1030+074
as here the lensed image is very bright in the optical. However, we
argue that for significantly fainter radio sources, the likelihood of
this happening is much less.}
\end{figure*}

\subsection{The method applied to the FIRST and SDSS surveys}

Currently, the large-area surveys with the highest resolution are the optical SDSS 
(York et al. 2000) which covers an area of 7000 square degrees of the North
Galactic Cap at an angular resolution of $\sim1\farcs4$, and the FIRST
radio survey at 1.4~GHz (Becker, White \& Helfand 1995) which covers a similar area
at a resolution of 5$^{\prime\prime}$. Both surveys claim astrometric
accuracy of a small fraction of an arcsecond. Neither survey has the
resolution unambiguously to distinguish image splittings in average
galaxy-lens systems of about 1$^{\prime\prime}$. However, using a test
based on the optical--radio offset and the misalignment angle 
should in this case be able to detect lens
systems with separation of less than 1$^{\prime\prime}$, given the
astrometric accuracy claimed by both surveys and the resolution of the
FIRST survey of slightly less than 5$^{\prime\prime}$. In principle,
the test can be made more efficient for double-image lens systems by
selecting misalignment angles close to 0$^{\circ}$ and for four-image
systems by selecting angles close to 90$^{\circ}$.

A few JVAS/CLASS lenses would not be detectable on the SDSS. The SDSS
flux limit is $r^{\prime}$=23.1 (York et al. 2000) which corresponds to
$V\sim 22.6$ (Jester et al. 2005), or to $I\sim 20.2$ given a typical
$V-I$ colour of an elliptical galaxy at $z\sim 0.5$ (Bruzual 1983). Lens
galaxy magnitudes are available for all JVAS/CLASS lenses, and six are
well below this limit: B0128+437 (Biggs et al. 2004), B0739+366 (Marlow
et al. 2001), B1127+385 (Koopmans et al. 1999), B1359+154 (Rusin et al.
2001) B1555+375 (Marlow et al. 1999) and B1938+666 (King et al. 1998).
B1555+375 is in the SDSS region, and indeed is not detected. Therefore
there is little prospect of using the methods outlined here for the
detection of about 30\% of radio source lenses.

There are other useful radio surveys which can be used to establish
spectral index of radio sources, which indicates whether the source is
extended (steep-spectrum) or compact (flat-spectrum), in particular the
Westerbork Northern Sky Survey (Rengelink et al. 1997) at 325MHz 
and the GB6 survey (Condon et al. 1998) at 5GHz. 

\section{A test of the survey concept}

\subsection{Sample selection}

Sources were selected from the regions of overlap between Data Release 3
of the SDSS and the FIRST survey. To avoid selection of radio lobes or
more extended radio sources, for which the procedure outlined in
section 2 
probably does not work, only sources with fitted deconvolved sizes
of 2\farcs5 or less from the FIRST survey were selected, and 3333 sources
already observed by the CLASS survey were also rejected. An additional 
spectral index selection was achieved by requiring that all selected 
sources have FIRST integrated flux densities $\geq$9~mJy and 
have spectral index flatter than $-0.5$ ($S_\nu\propto\nu^{\alpha}$) 
between the 1.4-GHz FIRST survey and detections in 
the 325-MHz Westerbork Northern 
Sky Survey. The spectral index selection allows one to reject small 
compact steep spectrum sources with twin radio lobes lying at different
distances from the radio core which would otherwise
contaminate the sample to an unacceptable extent. 
This produces a sample of 3363 sources, of which 1831
have a potential SDSS identification within
6$^{\prime\prime}$. A further cut was performed based on the SDSS
photometry in order to remove low-redshift $z<0.2$ galaxies which are
unlikely to be lenses by requiring that the fluxes in the SDSS $u$, $g$
and $z$ bands obey the relation $u-g<g-z$. This in effect discriminates
against galaxies which have a 4000-\AA\ break between the $u$ and $g$
bands. 1380 objects remain after applying this cut.

In Figure 3 we plot the optical--radio offset versus misalignment angle
$M$ for the sample of 1380 candidates. In addition, we have simulated
FIRST observations of the gravitational lenses from CLASS, reducing
their total flux density to 20~mJy and using their simulated FIRST
characteristics, together with the known lens position, to calculate the
optical--radio offset and misalignment angle $M$. Note that this assumes that the
optical emission in the simulated lenses comes only from the lens
galaxy, which is not true for the CLASS sample but is justified here because
the lenses we are trying to find come from the $<$100~mJy population in
which the sources are not optically bright.

\begin{figure}
\psfig{figure=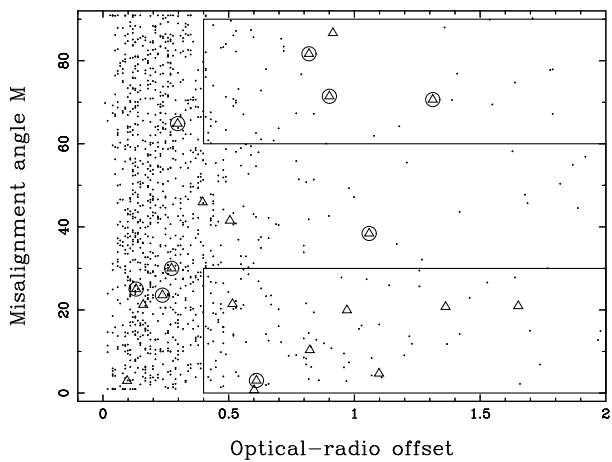,width=8cm,angle=-90}
\caption{Misalignment angle M vs. optical--radio offset, in arcsec, for
sources in the FIRST-SDSS overlap region  (specks). The CLASS lenses have
been rescaled to 20mJy and mapped using artificial VLA snapshots and
the same noise levels as FIRST. Double lenses  (triangles) lie mostly in a region of the
M:optical--radio offset plane well away from the bulk of the normal source population,
as do the quad lenses  (circled triangles). The exceptions are the small-separation
lenses which cannot be detected in this survey. The boxes enclose
regions selected for observation, which contain a much larger fraction
of gravitational lenses than the general population.}
\end{figure}

A final observing sample of 160 sources has been constructed by
choosing all objects with optical--radio offset between 0\farcs4 and
2$^{\prime\prime}$, and with misalignment angles
$0^{\circ}<M<30^{\circ}$ and $60^{\circ}<M<90^{\circ}$. This cuts out
the vast majority of unlensed sources, while keeping significant
numbers of lenses (12 out of 22 of the simulated lens population
remain in the final sample). Of these 12, we would expect $\sim$9 to have
lens galaxies detectable with the SDSS. This implies that in
observations of this sample we should find a fraction 9/22 of
lenses, while carrying out 160 observations instead of 3363, an
increase in efficiency of nearly a factor 10.  More lenses could be
discovered at only slightly lower efficiency by including the
$30^{\circ}<M<60^{\circ}$ range of misalignment angle.  This means
that in principle one lens could be discovered per 60 sources instead
of per 600, resulting in a lensing rate from an EVLA observation of 1
lens per 2 hours instead of the CLASS rate which was more like one
lens per day from the VLA\footnote{We note that if we had considered
sources selected by the method outlined here, but which {\em had} previously
been observed with CLASS, we would have ended up with an observing list
of 69 sources and ``discovered'' one lens, CLASS B1600+434. The other lens
in the appropriate region of sky, CLASS B1555+375, would have just been
missed due to its lensing galaxy lying below the SDSS threshold}. 
This factor of $\sim$10 improvement is of the
required order for the future extraction of 10 times larger samples of
lenses.


\section{Observations and results}

\subsection{Observations}

\subsubsection{VLA}

A total of 4.5 hours of VLA time was allocated to the pilot project in order
to test the survey concept. 
Sources were randomly chosen from this sample according to available
slots in the observing schedule. Three epochs of observations were
obtained, 2005 March 03 (31 sources), 2005 March 17 (35 sources) 
and 2005 March 29 (35 sources) for between 110 and 120 seconds per
source. The VLA in B-configuration was used at a frequency of 
8.415~GHz and a total bandwidth of 100~MHz (2 IFs each of 50~MHz 
bandwidth). This arrangement gives a maximum baseline of 12~km, a 
nominal resolution of $\sim0\farcs7$ and an rms noise
level of approximately 150~$\mu$Jy$\,$beam$^{-1}$. Phase calibrators
chosen from the JVAS survey were observed approximately every 20
minutes. These phase calibrators have a nominal astrometric accuracy of
40~mas or better. The phase stability throughout the observations was good,
typically with atmospheric phase drifts of $<30^{\circ}$ over the whole
observing session, and the accuracy of positions in the derived maps
should therefore be better than about 100~mas.

Unfortunately, the epoch from 2005 March 03 was affected by apparently
random phase jumps between the Stokes R and L channels, mainly in the
data on the target sources. Although correlated phases are obviously
present, standard calibration and imaging procedures were unable to
recover images of adequate quality for this epoch. The phases could be
lined up by initial self-calibration to a point-source model, but at
the risk of imposing spurious structure on the data and this epoch was
therefore not used in the subsequent analysis. The total observed
sample was therefore only 70 sources, and the expected yield was,
therefore, one gravitational lens system.

The two good epochs were processed in a standard way with the NRAO
Astronomical Image Processing System ({\sc aips}). The flux density scale was
normalised to an observation of 3C286 in each case (Baars et al. 1977)
and an initial phase solution was constructed by interpolation across
the phase calibrators. The amplitude and phase-calibrated data were
mapped by hand with the {\sc difmap} package (Shepherd 1997) including
CLEANing and one iteration of phase selfcalibration for sources stronger
than $\sim$10~mJy. Final maps have been produced using uniform weighting
and a restoring beam of 0\farcs5, slightly below the nominal resolution of
0\farcs67.

\subsubsection{MERLIN}

Two sources whose structures were ambiguous in the VLA observations,
J1121+444 and J1316+575, were observed with the Multi-Element
Radio-Linked Interferometer (MERLIN, Thomasson 1986) at an observing
frequency of 4.994~GHz, with a 32-MHz bandwidth and a resolution of
50~mas. Observations were obtained on 2006 July 14-16 for 32h
(J1316+575) and 17h (J1121+444) using four telescopes of the array
(Jodrell Mk2, Defford, Knockin and Cambridge). Phase calibrators were
observed for 2.5 minutes in every 10 minutes, these being J1307+562 and
J1115+416 for the target sources J1316+575 and J1121+444. The
point-source calibrator OQ208 and flux calibrator 3C286 were also
observed, for 75 and 27 minutes respectively. Data were
processed, and amplitude calibration applied, in the 
standard manner using the {\sc d-programme} software
available from the MERLIN site, and the flux density scale was set using the
Baars et al. (1977) flux density for 3C286 of 7.3~Jy at 5~GHz, reduced by 30\% to
allow for resolution effects on the shortest available baseline
(Knockin-Mk2). This reduction gives a flux density of 2.42~Jy for OQ208.
Phase solutions were produced in {\sc aips} for each scan of the two phase
calibrators and interpolated to produce an overall atmospheric phase 
calibration. Maps were then made, using natural weighting in {\sc aips
imagr} and without self-calibration steps, yielding images with an rms 
noise of $\sim$190~$\mu$Jy/beam for J1121+444 and
$\sim$110~$\mu$Jy/beam for J1316+575. This process produces clear detections of the 
phase calibrators. However, the maps of the two target sources do not show
any detectable radio emission.

\subsection{Notes on resolved sources}

Maps of the resolved objects are presented in Figures 4 and 5, and we discuss
each case in turn.

{\bf J1055+568.} This source has a steep spectrum between 1.4 and
8.4~GHz, declining from a peak brightness of 12mJy/beam in the FIRST survey
to a peak of 1.3mJy/beam in these maps. This is almost certainly because
at least one of the components is heavily resolved at 8.4~GHz. 
Assuming the SDSS astrometry to be correct, a gravitational
lens hypothesis is difficult to sustain as both radio components are on
the same side of the optical ID, and the stronger component is close to
the SDSS position. This source is almost certainly a steep-spectrum
source with a radio jet.

{\bf J1121+444.} This source is still a candidate gravitational lens
from the VLA observations. It
is not a very good candidate, as the FIRST position is relatively far
from the brighter 8.4-GHz component, suggesting a spectral index
gradient across the source which is inconsistent with lensing. This
impression is confirmed by the MERLIN map (Fig. 5), which resolves out all radio
emission. We conclude that both components are therefore extended on
scales larger than the MERLIN resolution of 50 mas, and this is
therefore not a gravitationally lensed compact source.

{\bf J1220+476.} The source is quite complex and appears to have a third
component which is just detected in the radio map. Again, the components
are almost certainly heavily resolved and the stronger radio component
is close to the SDSS galaxy. It is thus difficult to claim this as a
serious candidate for gravitational lensing.

{\bf J1239+447.} At first glance this object looks promising,
consisting of two apparently unresolved sources, one of which lies on
the FIRST identification and one on the SDSS. The SDSS identification
is with a $z=2.04$ object of $r$ magnitude 19.8, and the overall radio
spectrum is relatively flat ($S^{1.4}_{8.4}=0.48$). A lensing
identification is however most unlikely, for the following reason. The
fact that the FIRST centroid lies on one of the 8.4-GHz components,
both of which have almost equal brightness, implies that the two
components have different spectral indices. Such a difference
automatically rules out the two components as images in a
gravitational lens system. Moreover, the SDSS quasar coincides with
one 8.4~GHz component suggesting that the likely explanation for this
system is that it is a one-sided jet in a radio quasar, where the core
of the quasar is coincident with the SDSS identification and the
steep-spectrum radio jet is shifting the FIRST centroid of emission.

{\bf J1259+519.} As with J1239+447, this object is almost certainly a
source with a one-sided jet. The FIRST identification lies
almost on top of one 8.4-GHz component, and the SDSS identification (a
stellar object with $r$=21.5) on the other. In this system the overall
spectral index is very steep, suggesting that at least one component
is a heavily resolved radio lobe.

{\bf J1316+575.} This source is still a candidate from the VLA map, although the
displacement of the 8.4-GHz component from the FIRST position and the
position of the SDSS identification close to the brighter 8.4-GHz
component make it an unlikely one. Again, however, the MERLIN map (Fig.
5) resolves out both components and this is therefore not a
gravitationally lensed compact source.

{\bf J1321+476.} The brighter component in the 8.4-GHz map is
obviously resolved. In principle this might be thought to be the
merging two images of a 4-image lens system. However, the SDSS
identification is in the wrong place either for a lensing galaxy or to
be identified as a lensed radio image. Again, this system almost
certainly consists of a compact core coincident with the
SDSS position and a steep-spectrum radio jet which appears in FIRST.

{\bf J1325+540.} This radio source has a marginally detected secondary,
but both components are very steep-spectrum and heavily resolved.

{\bf J1441+567.} This is potentially a good lens candidate. Unlike in
the other cases, the FIRST position is between the two 8.4-GHz
components detected on our radio map, suggesting that the two 8.4-GHz
components have the same radio spectrum. However, the source has a
relatively steep radio spectrum and there is a faint bridge of radio emission
connecting the two components. This bridge is inconsistent with the
results of 2-image lensing and entirely consistent it being  a
twin-jetted double radio source, whose radio emission is not quite
steep-spectrum enough to cause the source to be removed by the
spectral index criterion.

The conclusion is that no clear examples of gravitational lensing have been
discovered in these observations.

\begin{figure*}
\begin{tabular}{ccc}
\psfig{figure=P1055+568,width=5.5cm,angle=-90}&
\psfig{figure=P1121+444,width=5.5cm,angle=-90}&
\psfig{figure=P1220+476,width=5.5cm,angle=-90}\\
\psfig{figure=P1239+447,width=5.5cm,angle=-90}&
\psfig{figure=P1259+519,width=5.5cm,angle=-90}&
\psfig{figure=P1316+575,width=5.5cm,angle=-90}\\
\psfig{figure=P1321+476,width=5.5cm,angle=-90}&
\psfig{figure=P1325+540,width=5.5cm,angle=-90}&
\psfig{figure=P1441+567,width=5.5cm,angle=-90}\\
\end{tabular}
\caption{Resolved sources from the VLA observations. The FIRST and SDSS
positions are indicated by five-pointed stars and circles respectively.
Top row, left to right: J1055+568, J1121+444, J1220+476. Middle row,
left to right: J1239+447, J1259+519, J1316+575. Bottom row, left to
right: J1321+476, J1325+540, J1441+567.}
\end{figure*}

\begin{figure*}
\begin{tabular}{cc}
\psfig{figure=M1118+447,width=6.5cm,angle=-90}&
\psfig{figure=M1315+578,width=6.5cm,angle=-90}\\
\end{tabular}
\caption{The two possible candidates from the VLA observations, observed
with the MERLIN array at 5GHz. No detection is made, and these sources
do not therefore contain compact components. Symbols are given as for
Figure 4.}
\end{figure*}

\subsection{Nature of the observed sources}

In the last section we considered the sources in which our VLA
observations revealed resolved structure. 
We now consider the sources that are unresolved by the 8.4-GHz VLA
observations. Figure 6 shows the position offsets determined between our
8.4-GHz radio component and the FIRST and SDSS surveys. It can be seen
that in the majority of the anomalous cases, the problem is an offset
between the VLA position and the SDSS identification.

\begin{figure}
\psfig{figure=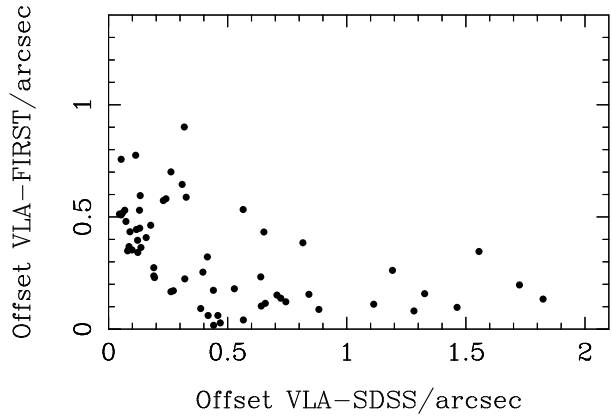,width=8cm,angle=-90}
\caption{Position offsets for unresolved sources between our VLA
measurements and the positions recorded in the FIRST and SDSS catalogues.}
\end{figure}

The SDSS source density is approximately 0.0021 per square arcsecond.
Hence in a search of 3764 sources within a radius of
2$^{\prime\prime}$ we would expect 98$\pm$10 chance
identifications. By reducing the sample of SDSS IDs from 2033 to 1538
using the colour cut, we reduce the expected number of chance
identifications to 74, and the selection on $M$ and optical--radio
offset reduces this further to 48$\pm$7 within the final sample of 160
(of which 70 were actually observed).  Therefore about 30\% of
associations between FIRST sources and SDSS objects are likely to be
chance coincidences, with the probability of a chance coincidence
increasing as the SDSS object goes further from the radio source. The
remaining 70\% are most probably cases in which there is an
astrophysical reason for a optical--radio offset such as a core-jet
structure in a radio quasar, cases in which the tail of the error
distribution of FIRST centroids reaches $\sim3\sigma$, or where the
SDSS astrometry is in error. This is entirely consistent with the
distribution in Figure 6, which shows that the distribution of offsets
between VLA and SDSS positions has a long tail, consisting of about
one-third of the observed objects which are probably chance
coincidences between FIRST and SDSS objects.

\section{Conclusion}

We have shown that it is in principle possible to increase the
efficiency of lens searches in compact radio sources by up to a factor
of $\sim$10, resulting in a lensing rate of 1 in 60 rather than 1 in 600
in the most favourable case. This can be achieved by using sub-resolution
position shifts and slight resolution effects within the radio FIRST
survey and the SDSS survey, supplemented by radio observations.
We present a preliminary programme with the VLA and MERLIN in which 
we would have expected to detect 1 lens out of an observed
sample of 70; we have eliminated all candidates and therefore detect no
lenses. 

The major difficulty with this method is that its efficiency is
fundamentally limited by the high density of optical objects at the SDSS
$r^{\prime}=23.1$, which means that each true lens system will be
accompanied by several tens of false-positives from random coincidences with
unrelated SDSS objects. In principle, this could be reduced further by
more stringent colour cuts to eliminate SDSS identifications which are
unlikely lensing galaxies. Further false-positives are provided by the
tail of sources with 3-$\sigma$ errors in the optical astrometry, and
efficiency would be increased dramatically by a 30\% reduction in
astrometric errors. The final problem is similar to the limiting
factor in the CLASS survey, namely that one runs out of radio sources
even at a lower flux level. In the current survey we have observed a third of the
available radio sources with $S_{\rm1.4GHz}>9\,$mJy in the SDSS footprint.
Further progress would require a larger volume of sky observed by SDSS,
or a deeper radio survey at a frequency sufficiently different from
1.4~GHz to allow spectral selection. Alternatively, one could relax the
spectral index criterion in the sample selection, at a cost to
efficiency.

A major advantage with the method, however, is that lens systems are
radio loud and can be confirmed by purely radio observations rather
than requiring high-resolution optical observations using scarce
resources such as the Hubble Space Telescope.  Moreover, in the future
this method may find more widespread applicability as all-sky surveys
with astrometric precisions of 0.1 arcseconds or better become
commonplace. Such surveys are likely to be done in the next ten years
with, for example, the Large Synoptic Survey Telescope (LSST, Claver
et al.  2004) in the optical, and in the radio with the Low-Frequency
Array Radiotelescope (LOFAR, R\"ottgering et al. 2005).  These surveys
should give rise to samples of many millions of radio sources with
optical identifications and spectral index information, allowing the
efficient discovery of thousands of lenses with modest followup time.

\section*{Acknowledgements}
The National Radio Astronomy Observatory is a facility of the National
Science Foundation operated under cooperative agreement by Associated
Universities, Inc. The MERLIN array is a
national facility operated by the University of Manchester at
Jodrell Bank Observatory on behalf of PPARC.
We thank all involved in the FIRST survey for
providing this resource. 
Funding for the Sloan Digital Sky Survey (SDSS) has been provided by
the Alfred P. Sloan Foundation, the Participating Institutions, the
National Aeronautics and Space Administration, the National Science
Foundation, the U.S. Department of Energy, the Japanese Monbukagakusho,
and the Max Planck Society. The SDSS Web site is http://www.sdss.org/.
The SDSS is managed by the Astrophysical Research Consortium (ARC)
for the Participating Institutions. The Participating Institutions are
The University of Chicago, Fermilab, the Institute for Advanced Study,
the Japan Participation Group, The Johns Hopkins University, the Korean
Scientist Group, Los Alamos National Laboratory, the
Max-Planck-Institute for Astronomy (MPIA), the Max-Planck-Institute for
Astrophysics (MPA), New Mexico State University, University of
Pittsburgh, University of Portsmouth, Princeton University, the United
States Naval Observatory, and the University of Washington.
This research has made use of the NASA/IPAC
Extragalactic Database (NED) which is operated by the Jet Propulsion
Laboratory, California Institute of Technology, under contract with
the National Aeronautics and Space Administration.  This work was
supported in part by the European Community's Sixth Framework Marie
Curie Research Training Network Programme, Contract No.
MRTN-CT-2004-505183 "ANGLES".

\section*{References}

\noindent Baars J.W.M., Genzel R., Pauliny-Toth I.I.K., Witzel A., 1977,  A\&A, 61, 99. 

\noindent Bahcall J.N., Maoz D., Doxsey R., Schneider D.P., Bahcall N.A., Lahav O., Yanny B., 1992,  ApJ, 387, 56. 

\noindent Becker R.H., White R.L., Helfand D.J., 1995,  ApJ, 450, 559. 

\noindent Bennett C.L., Lawrence C.R., Burke B.F., Hewitt J.N., Mahoney J., 1986,  ApJS, 61, 1. 

\noindent Biggs A.D., Browne I.W.A., Jackson N.J., York T., Norbury M.A., McKean J.P., Phillips P.M., 2004,  MNRAS, 350, 949. 

\noindent Bolton A.S., Burles S.,  Koopmans L.V.E., Treu T., Moustakas, L.A., 2006,  ApJ, 638, 703. 

\noindent Boyce E., Winn J.N., Hewitt J.N., Myers S.T., 2006, ApJ, 648, 73.

\noindent Browne I.W.A., Wilkinson P.N., Jackson N.J.F., Myers S.T., Fassnacht C.D., Koopmans, L.V.E., Marlow D.R., Norbury M., Rusin D., Sykes C.M.,  2003,  MNRAS, 341, 13. 

\noindent Bruzual A.G., 1983,  ApJS, 53, 497. 

\noindent Claver C.F., et al., 2004, SPIE 5489, 705

\noindent Cohn J.D., Kochanek C.S., McLeod B.A., Keeton C.R., 2001,  ApJ, 554, 1216. 

\noindent Condon J.J., Cotton, W.D., Greisen E.W., Yin Q.F., Perley R.A., Taylor G.B., Broderick J.J., 1998,  AJ, 115, 1693. 

\noindent Crampton D., McClure R.D., Fletcher J.M., 1992,  ApJ, 392, 23. 

\noindent Dye S., Warren S.J., 2005,  ApJ, 623, 31. 

\noindent Falco E.E., Kochanek C.S., Munoz J.A., 1998,  ApJ, 494, 47. 

\noindent Haarsma D.B., Winn J.N., Falco E.E., Kochanek C.S., Ammar P., Boersma C., Fogwell S., Muxlow T.W.B., McLeod B.A., Leh\'ar J., 2005,  AJ, 130, 1977. 

\noindent Hewitt J.N., Turner E.L., Schneider D.P., Burke B.F., Langston G.I., 1988,  Natur, 333, 537. 

\noindent Impey C.D., Falco E.E., Kochanek C.S., Leh\'ar J., McLeod B.A., Rix, H.-W., Peng C.Y., Keeton C.R., 1998,  ApJ, 509, 551. 

\noindent Inada N., et al., 2003,  Nature, 426, 810

\noindent Jaunsen A.O., Jablonski M., Pettersen B.R., Stabell R., 1995,  A\&A, 300, 323. 

\noindent Jester S., Schneider D.P., Richards G.T., Green R.F., Schmidt M., Hall P.B., Strauss M.A., VandenBerk D.E., Stoughton C., Gunn J.E.,  2005,  AJ, 130, 873. 

\noindent King L.J., Browne I.W.A., Marlow D.R., Patnaik A.R., Wilkinson P.N., 1999,  MNRAS, 307, 225. 

\noindent King L.J., Jackson N., Blandford R.D., Bremer M.N., Browne I.W.A., deBruyn A.G., Fassnacht C., Koopmans L., Marlow D., Wilkinson P.N., 1998,  MNRAS, 295L, 41. 

\noindent Kochanek C.S., Falco E.E., Schild R., 1995,  ApJ, 452, 109. 

\noindent Kochanek C.S., Keeton C.R., McLeod B.A., 2001,  ApJ, 547, 50. 

\noindent Kochanek C.S., 2002,  ApJ, 578, 25. 

\noindent Koopmans L.V.E., Treu T., Fassnacht C.D., Blandford R.D., Surpi G., 2003,  ApJ, 599, 70. 

\noindent Koopmans L.V.E., de Bruyn A.G., Marlow D.R., Jackson N., Blandford R.D., Browne I.W.A., Fassnacht C.D., Myers S.T., Pearson T.J., Readhead A.C.S.,  1999,  MNRAS, 303, 727. 

\noindent Koopmans L.V.E., et al., astro-ph/0601628

\noindent Lawrence C.R., Bennett C.L., Hewitt J.N., Langston G.I., Klotz S.E., Burke B.F., Turner K.C., 1986,  ApJS, 61, 105. 

\noindent Leh\'ar J., Buchalter A., McMahon R.G., Kochanek C.S., Muxlow T.W.B., 2001,  ApJ, 547, 60. 

\noindent Maoz D., Bahcall J.N., Doxsey R., Schneider D.P., Bahcall N.A., Lahav O., Yanny B., 1992,  ApJ, 394, 51. 

\noindent Marlow D.R., Rusin D., Jackson N., Wilkinson P.N., Browne I.W.A., Koopmans L., 2000,  AJ, 119, 2629. 

\noindent Marlow D.R., Myers S.T., Rusin D., Jackson N., Browne I.W.A., Wilkinson P.N., Muxlow T., Fassnacht C.D., Lubin L., Kundic T.,  1999,  AJ, 118, 654. 

\noindent Marlow D.R., Rusin D., Norbury M., Jackson N., Browne I.W.A., Wilkinson P.N., Fassnacht C.D., Myers S.T., Koopmans L.V.E., Blandford R.D.,  2001,  AJ, 121, 619. 

\noindent Mu\~noz J.A., Falco E.E., Kochanek C.S., Leh\'ar J., Mediavilla E., 2003,  ApJ, 594, 684. 

\noindent Myers S.T., Jackson N.J., Browne I.W.A., deBruyn A.G., Pearson T.J., Readhead A.C.S., Wilkinson P.N., Biggs A.D., Blandford R.D., Fassnacht C.D.,  2003,  MNRAS, 341, 1. 

\noindent Oguri M., et al., 2006, astro-ph/0605571

\noindent Patnaik A.R., Browne I.W.A., Wilkinson P.N., Wrobel J.M., 1992,  MNRAS, 254, 655. 

\noindent Pindor B., Turner E.L., Lupton R.H., Brinkmann J., 2003, AJ, 125, 2325

\noindent Refsdal S. 1964,  MNRAS, 128, 295. 

\noindent Rengelink R.B., Tang Y., de Bruyn A.G., Miley G.K., Bremer M.N., R\"ottgering H.J.A., Bremer M.A.R., 1997.
A\&AS, 124, 259

\noindent R\"ottgering H., van Haarlem M., Miley G., 
In ``Probing Galaxies through Quasar Absorption Lines'', IAU Colloquium, 
Proc. IAU 199, held March 14-18 2005, Shanghai, ed. Williams P.R. et al.,
Cambridge University Press 2005. P.381

\noindent Rusin D., Kochanek C.S., Norbury M., Falco E.E., Impey C.D., Leh\'a¡r J., McLeod B. A., Rix, H.-W., Keeton C.
R. ,Mu\~n±oz J.A., Peng Y., 2001, ApJ, 557, 594

\noindent Saha P., Williams L.L.R., 2001,  AJ, 122, 585. 

\noindent Schechter P.L., Bailyn C.D., Barr R., Barvainis R., Becker C.M., Bernstein G.M.,
Blakeslee J.P., Bus S.J., Dressler A., Falco E.E.,  1997,  ApJ, 475L, 85. 

\noindent Shepherd M., in Astronomical Data Analysis Software and Systems VI A.S.P. Conference
Series, Vol. 125, 1997, Gareth Hunt and H. E. Payne, eds. P. 77.

\noindent Surdej J., Claeskens J.F., Crampton D., Filippenko A.V., Hutsemekers D., Magain P.,
Pirenne B., Vanderriest C., Yee, H.K.C., 1993,  AJ, 105, 2064. 

\noindent Thomasson P., 1986, QJRAS, 27, 413

\noindent Treu T., Koopmans L.V.E., 2004,  ApJ, 611, 739. 

\noindent Treu T., Koopmans L.V.E., Bolton A.S., Burles S., Moustakas L.A., 2006,  ApJ, 640, 662. 

\noindent Walsh D., Carswell R.F., Weymann R.J. 1979,  Nat, 279, 381. 

\noindent Willis J.P., Hewett P.C., Warren S.J., Dye S., Maddox N., astro-ph/0603421

\noindent Winn J.N., Hewitt J.N., Patnaik A.R., Schechter P.L., Schommer R.A., L\'opez S., Maza J., Wachter S., 2001,  AJ, 121, 1223. 

\noindent Winn J.N., Morgan N.D., Hewitt J.N., Kochanek C.S., Lovell J.E.J., Patnaik A.R., Pindor B., Schechter P.L., Schommer R.A., 2002,  AJ, 123, 10. 

\noindent York D.G., et al., 2000,  AJ, 120, 1579.

\end{document}